\documentclass[aps, prd, twocolumn, print, nofootinbib]{revtex4}
\usepackage{amsmath}
\usepackage{txfonts}
\usepackage{amssymb}
\usepackage{mathrsfs}
\usepackage{graphicx}
\usepackage{epsfig}
\usepackage{dcolumn}
\usepackage{bm}
\usepackage{ulem}
\usepackage{color}

\begin{document}


\title{Constraints from compact star observations  on
  non-Newtonian gravity in strange stars based on a
 density dependent quark mass model}

\author{Shu-Hua Yang$^{1}$}\email{ysh@mail.ccnu.edu.cn}
\author{Chun-Mei PI$^{2}$} \author{Xiao-Ping Zheng$^1$$^,$$^3$}
\author{Fridolin Weber$^4$$^,$$^5$}

\affiliation{$^1$Institute of Astrophysics, Central China Normal
  University, Wuhan 430079, China\\$^2$School of Physics and
  Mechanical \& Electrical Engineering, Hubei University of Education,
  Wuhan 430205, China \\$^3$Department of Astronomy, School of
  physics, Huazhong University of Science and Technology, Wuhan
  430074, China \\$^4$Department of Physics, San Diego State
  University, San Diego, CA 92182, USA \\$^5$ Center for Astrophysics
  and Space Sciences, University of California at San Diego, La Jolla,
  CA 92093, USA}

\date{November 2020}

\begin{abstract}
Using a density dependent quark mass (QMDD) model for strange quark
matter, we investigate the effects of non-Newtonian gravity on the
properties of strange stars and constrain the parameters of the QMDD
model by employing the mass of PSR J0740+6620 and the tidal
deformability of GW170817. We find that for QMDD model these
mass and tidal deformability observations would rule out the existence
of strange stars if non-Newtonian gravity effects are ignored. For
the current quark masses of $m_{u0}=2.16$ MeV,
$m_{d0}=4.67$ MeV, and $m_{s0}=93$ MeV, we find that a strange star
can exist for values of the non-Newtonian gravity parameter
$g^{2}/\mu^{2}$ in the range of 4.58 GeV$^{-2}\leq g^{2}/\mu^{2}\leq$
9.32 GeV$^{-2}$, and that the parameters $D$ and $C$ of the QMDD model
are restricted to 158.3 MeV$\leq D^{1/2}\leq$ 181.2 MeV and $-0.65
\leq C \leq -0.12$. It is found that the largest
possible maximum mass of a strange star obtained with the QMDD
model is $2.42 \, M_{\odot}$, and that the secondary component
of GW190814 with a mass of $2.59_{-0.09}^{+0.08}\,
M_{\odot}$ could not be a static strange star. We also find that for
the mass and radius of PSR J0030+0451 given by Riley et al.\ through
the analysis of observational data of NICER, there exists a very tiny
allowed parameter space for which strange stars computed for the QMDD
model agree with the observations of PSR J0740+6620, GW170817 and PSR
J0030+0451 simultaneously. However, for the mass and radius given by
Miller et al., no such parameter space exist.
\end{abstract}

\maketitle

\section{INTRODUCTION} \label{S:intro}

As hypothesized by Itoh \citep{itoh70}, Bodmer \citep{bod71}, Witten
\citep{wit84}, and Terazawa \citep{tera89}, strange quark matter
(SQM) consisting of up ($u$), down ($d$) and strange ($s$) quarks and
electrons may be the true ground state of baryonic matter. According
to this hypothesis, compact stars made entirely of SQM, referred to as
strange stars (SSs), ought to exist in the universe
\cite{far84,alc86,hae86,alc88,mad99,web05}.

Effects of non-Newtonian gravity on the properties of neutron stars
and SSs have been studied extensively
\citep[e.g.,][]{kri09,wen09,sul11,zha11,yan13,lin14,lu17,yu18,yang20}. The
conventional inverse-square-law of gravity is expected to be violated
in the efforts of trying to unify gravity with the other three
fundamental forces, namely, the electromagnetic, weak and strong
interactions \citep{fis99,ade03,ade09}. Non-Newtonian gravity arise
due to either the geometrical effect of the extra space-time
dimensions predicted by string theory and/or the exchange of weakly
interacting bosons, such as a neutral very weakly coupled spin-1 gauge
U-boson proposed in the super-symmetric extension of the standard
model \citep{Fay80,Fay81}. Although the existence of non-Newtonian
gravity is not confirmed yet, constraints on the upper limits of the
deviations from Newton's gravity have been set experimentally (see
\cite{mur15} and references therein).

For the standard MIT bag model, Yang et al. \cite{yang20} found that
if non-Newtonian gravity effects are ignored, the existence of SSs is
ruled out by the mass of PSR J0740+6620 ($2.14_{-0.09}^{+0.10}\,
M_{\odot}$ for a 68.3\% credibility interval; $2.14_{-0.18}^{+0.20}\,
M_{\odot}$ for a 95.4\% credibility interval) \cite{cro20} and the
dimensionless tidal deformability of a $1.4\, M_{\odot}$ star of
GW170817 ($\Lambda(1.4)=190 _{-120}^{+390}$)
\cite{abb17,abb18}. However, if non-Newtonian gravity effects are
considered, Yang et al. \cite{yang20} found that SSs can exist for
certain ranges of the values of the non-Newtonian gravity parameter
$g^{2}/\mu^{2}$, and the bag constant $B$ and the strong interaction
coupling constant $\alpha_{S}$ of the SQM model. For example, for a
strange quark mass of $m_{s}=95$ MeV, SSs can exist for 1.37
GeV$^{-2}\leq g^{2}/\mu^{2}\leq$ 7.28 GeV$^{-2}$, and limits on
parameters of the SQM model are 141.3 MeV$\leq B^{1/4}\leq$ 150.9 MeV
and $\alpha_{S}\leq 0.56$.


Recently, the QMDD model was revisited in detail by Backes et
al. \cite{bac20} without the inclusion of the non-Newtonian
effects. Similar to the results given by Yang et al. \cite{yang20},
they found that the observations of GW170817 and the mass of PSR
J0740+6620 cannot be satisfied simultaneously for SSs with the QMDD
model. These authors did not use the constraints of the
dimensionless tidal deformability of a $1.4\, M_{\odot}$ star from
GW170817 directly. Instead, they employed the radius of a $1.4\,
M_{\odot}$ star, which is $R_{1.4} = 11.0_{-0.6}^{+0.9}$ km, derived
from the observations of GW170817 by Capano et al. \citep{cap20}.

In this paper, we will investigate the effects of non-Newtonian
gravity on the properties of SSs and constrain the parameter space of the QMDD model using the tidal deformability of GW170817 and the mass of PSR
J0740+6620. Moreover, constraints from the mass
and radius of PSR J0030+0451 derived from  NICER observations
\cite{ril19, mil19} are investigated too.

This paper is organized as follows. In Sec. II, we briefly review the
QMDD model and the equation of state (EOS) of SQM including the
non-Newtonian gravity effects. In Sec. III, numerical results and
discussions are presented. Finally, a brief summary of our results is
given in Sec. IV.

\section{EOS of SQM including the non-Newtonian gravity effects}\label{Sec II}

Before discussing the effects of non-Newtonian gravity on the EOS of
SQM, we briefly review the phenomenological model for the EOS employed
in this paper, namely the QMDD model.

The key feature of the QMDD model is the use of density dependent
quark masses to express non-perturbative interaction effects
\cite{fow81,plu84}. The first few QMDD studies of the EOS of SQM
were thermodynamicall inconsistent \citep[e.g.,][]{peng00,tor13,xia14}.  Furthermore, while the original
quark mass scaling formalism barely accounted for the
confinement interaction \citep[e.g.,][]{fow81,peng99}, an improved quark mass scaling taking into account
both the linear confinement and leading order interactions has been
introduced by Xia et al. \citep{xia14}.

Taking into account both the linear confinement and leading order
interactions, the quark mass scaling is given by \cite{xia14}
\begin{equation}
m_{i} = m_{i0} + m_I \equiv m_{i0} + \frac{D}{n_b^{1/3}} + Cn_{b}^{1/3}.
\label{masses}
\end{equation}
Here $m_I$ is a density dependent term that includes the quark
interaction effects introduced through the adjustable parameters $C$
and $D$, $m_{i0}$ is the current mass of quark flavor $i$ with $m_{u0}
= 2.16$ MeV, $m_{d0} = 4.67$ MeV, and $m_{s0} = 93$ MeV
\citep{zyla20}, and $n_b$ is the baryonic density
\begin{equation}
n_{b} = \frac{1}{3} \sum_i n_i ,
\label{baryon-number-density}
\end{equation}
where the number density of each quark species $n_i$ is  given by
Eq. (\ref{number-density}).

The EOS of SQM with the above density dependent quark masses is to be
determined subject to the following fully consistent thermodynamic
conditions \cite{xia14}.
At zero temperature, the thermodynamic
potential of free unpaired particles is given by
\begin{equation}
\Omega_0 = -\sum_i \frac{g}{24 \pi^2} \left[ \mu_i^*\nu_i \left(
  \nu_i^2 - \frac{3}{2}m_i^2 \right) + \frac{3}{2} m_i^4 \ln
  \frac{\mu_i^* + \nu_i}{m_i} \right],
\label{thermodynamic-potential-free-system}
\end{equation}
where $g = 6$ is the degeneracy of quarks, $\mu_i^*$ is the effective
chemical potential of quark flavor $i$, and it is related to the
chemical potential $\mu_i$ through the following equation,
\begin{equation}
\mu_i = \mu_i^* + \frac{1}{3}\frac{\partial m_I}{\partial n_b}
\frac{\partial \Omega_0}{\partial m_I} .
\label{real-effective-chemical-potentials}
\end{equation}
The quantity $\nu_i$ denotes the Fermi momentum of a quark of type $i$,
\begin{equation}
\nu_i = \sqrt{\mu_i^{*2} - m_i^2}
\label{eq:fermimom}
\end{equation}
and the corresponding particle number
densities are given by
\begin{equation}
n_i = \frac{g}{6\pi^2}(\mu_i^{*2} - m_i^2)^{3/2} =
\frac{g\nu_i^3}{6\pi^2}.
\label{number-density}
\end{equation}
The energy density without the effects of the non-Newtonian gravity is
given by
\begin{equation}
\epsilon_{Q} = \Omega_0 - \sum_i \mu_i^*
\frac{\partial\Omega_0}{\partial\mu_i^*},
\label{eq:epsQ}
\end{equation}
and the pressure is obtained from
\begin{equation}
p_{Q} = -\Omega_0 + \sum_{i,j} \frac{\partial\Omega_0}{\partial m_j}
n_i \frac{\partial m_j}{\partial n_i},
\label{pressure}
\end{equation}
which can be written in the more convenient form
\begin{equation}
p_{Q} = -\Omega_{0}+n_{\mathrm{b}} \frac{\partial
  m_{\mathrm{I}}}{\partial n_{\mathrm{b}}} \frac{\partial
  \Omega_{0}}{\partial m_{\mathrm{I}}}.
\label{eq:pQ}
\end{equation}

In addition, chemical equilibrium is maintained by the
weak-interaction of SQM, which leads for the chemical potentials to
the following conditions,
\begin{eqnarray}
\mu_{d} &=&  \mu_{s} , \\
\mu_{s} &=& \mu_{u}+\mu_{e} .
\end{eqnarray}
The electric charge neutrality condition is given by
\begin{equation}
\frac{2}{3}n_{u}-\frac{1}{3}n_{d}-\frac{1}{3}n_{s}-n_{e}=0.
\end{equation}

Non-Newtonian gravity is often characterized effectively by adding a
Yukawa term to the normal gravitational potential
\citep{fuj71}.\footnote{An extra Yukawa term also naturally arises in
  the weak-field limit of some modified theories of gravity, e.g.,
  f(R) gravity, the nonsymmetric gravitational theory, and Modified
  Gravity. See \cite{li19}, and references therein.} The Yukawa-type
non-Newtonian gravity between the two objects with masses $m_{1}$ and
$m_{2}$ is \citep{fis99,ade03,ade09}
\begin{equation}
V(r)=-\frac{G_{\infty}m_{1}m_{2}}{r} \left( 1+\alpha e^{-r/\lambda}
\right) = V_N(r) + V_Y(r),
\label{vr}
\end{equation}
where $V_Y(r)$ is the Yukawa correction to the Newtonian potential
$V_N(r)$.  The quantity $G_{\infty} = 6.6710\times 10^{-11}~\rm{N} \,
{\rm m}^2/{\rm kg}^2$ is the universal gravitational constant,
$\alpha$ is the dimensionless coupling constant of the Yukawa force,
and $\lambda$ is the range of the Yukawa force mediated by the
exchange of bosons of mass $\mu$ (given in natural units) among $m_1$
and $m_2$,
\begin{equation}
\lambda=\frac{1}{\mu}.
\label{mulam}
\end{equation}
In this picture, the Yukawa term is the static limit of an interaction
mediated by virtual bosons.  The strength parameter in Eq.\ (\ref{vr})
is given by
\begin{equation}
\alpha=\pm \frac{g^{2}}{4\pi G_{\infty}m_{b}^{2}},
\end{equation}
where the $\pm$ sign refers to scalar (upper sign) or vector (lower
sign) bosons, $g$ is the boson-baryon coupling constant, and $m_{b}$
is the baryon mass.

Krivoruchenko et al. \cite{kri09} suggested that a neutral very weakly
coupled spin-1 gauge U-boson proposed in the super-symmetric extension
of the standard model is a favorite candidate for the exchanged boson
\citep{Fay80,Fay81}. This light and weakly interacting U-boson has
been used to explain the 511 keV $\gamma$-ray observation from the
galatic bulge \citep{Jean03,Boe04a,Boe04b}, and various experiments in
terrestrial laboratories have been proposed to search for this boson
\citep{Yong13}. Since the new bosons contribute to the EOS of dense
matter in terms of  $g^{2}/\mu^{2}$ \citep{Fuj88},
which can be large even when both the coupling constant $g$
and the mass $\mu$ of the light and weakly interacting bosons are
small, the structure of compact stars may be greatly influenced by the
non-Newtonian gravity effects.

It has been shown by Krivoruchenko et al. \cite{kri09} that an
increase of $g$ (a decrease of $\mu$) of scalar bosons has a negative
contribution to pressure, which makes the EOS of dense matter softer
and reduces the maximum mass of a compact star. By contrast, an
increase of $g$ (a decrease of $\mu$) of vector bosons makes the EOS
of dense matter stiffer and increases the maximum mass of a compact
star. In the following, we will only study the case of vector bosons
since a stiff EOS of SQM is needed to accomodate the tidal deformability of GW170817 and the mass of PSR
J0740+6620.

The contribution of the Yukawa correction $V_Y(r)$ of Eq.\ (\ref{vr})
to the energy density of SQM is obtained by integrating over the quark
densities $n_b(\vec x_1)$ and $n_b(\vec x_2)$ inside a
given volume $V$ \citep{Long03,kri09,wen09,lu17},
\begin{equation}
\epsilon_{Y}=\frac{1}{2V}\int 3n_{b}(\vec{x}_{1})\frac{g^{2}}{4\pi}
\frac{e^{-\mu r}}{r}3n_{b}(\vec{x}_{2})d\vec{x}_{1}d\vec{x}_{2} ,
\label{inte1}
\end{equation}
where $r = |\vec{x}_{1} - \vec{x}_{2}|$. The prefactors of 3 in front
of the quark densities are required since the baryon number of quarks
is $1/3$. Equation (\ref{inte1}) can be evaluated further since the
quark densities $n_{b}(\vec{x}_{1}) = n_{b}(\vec{x}_{2}) \equiv n_{b}$
are essentially independent of position
\citep{alc86,alc88,mad99,web05}. Moving $n_{b}$ outside of the
integral then leads for the energy density of SQM inside of $V=4\pi
R^3/3$ (for simplicity taken to be spherical\footnote{The actual
  geometry of the volume is unimportant since we are only interested
  in the local modification of the energy (Eq.\ (\ref{eq:epsY}))
  caused by the Yukawa term.}) to \citep{lu17,yang20}
\begin{equation}
\epsilon_{Y} = \frac{9}{2}g^{2}n_{b}^{2} \int_{0}^{R} r e^{-\mu r} dr.
\label{inte2}
\end{equation}
Upon carrying out the integration over the spherical volume one
arrives at
\begin{equation}
\epsilon_{Y} = \frac{9}{2} \frac{g^{2}n_{b}^{2}} {\mu^{2}}
\left[1-(1+\mu R) e^{-\mu R} \right] .
\label{ey1}
\end{equation}
Because the system we are considering is in principle very large, we
may take $R\rightarrow \infty$ in Eq.\ (\ref{ey1}) to arrive at
\begin{equation}
\epsilon_{Y} = \frac{9}{2} \frac{g^{2}}{\mu^{2}}n_{b}^{2} .
\label{eq:epsY}
\end{equation}
This analysis shows that the additional contribution to the energy
density from the Yukawa correction, $V_Y$, is simply determined (aside
from some constants) by the number of quarks per volume. The total
energy density of SQM is obtained by adding $\epsilon_Y$ to the
standard expression for the energy density of SQM given by
Eq.\ (\ref{eq:epsQ}), leading to
\begin{equation}
\epsilon = \epsilon_{Q} + \epsilon_{Y} .
\end{equation}
Correspondingly, the extra pressure due to the Yukawa correction is
\begin{equation}
p_{Y}=n_{b}^{2}\frac{d}{dn_{b}}\bigg(\frac{\epsilon_{Y}}{n_{b}}\bigg)
= \frac{9}{2}\frac{g^{2}n_{b}^{2}}{\mu^{2}}
\bigg(1-\frac{2n_{b}}{\mu}\frac{\partial \mu}{\partial n_{b}}\bigg).
\end{equation}
Assuming a constant boson mass (independent of the density)
\citep{kri09,wen09,lu17}, one obtains
\begin{equation}
p_{Y}=\epsilon_{Y} = \frac{9}{2} \frac{g^{2}}{\mu^{2}}n_{b}^{2}.
\end{equation}
The total pressure including the non-Newtonian gravity (Yukawa) term
then reads
\begin{equation}
p = p_{Q} + p_{Y} ,
\end{equation}
where $p_Q$ is given by Eq.\ (\ref{eq:pQ}).

\section{results and discussions}\label{Sec III}

For a given SQM EOS, the structure of strange stars and their tidal
deformability is calculated from the Tolman-Oppenheimer-Volkoff
equation, as described in
Refs. \citep{hin08,fla08,dam09,hin10,pos10,lat16}.

\begin{figure}
\resizebox{\hsize}{!}{\includegraphics{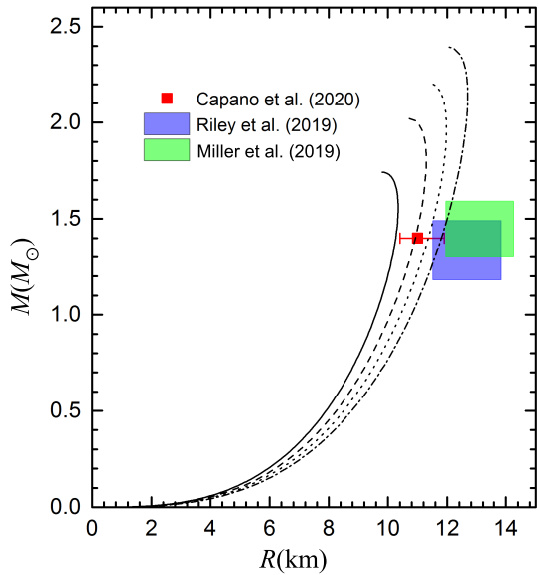}}
\caption{ (Color online) The mass-radius relation of SSs with $D^{1/2}
  = 161.3$ MeV, $C = -0.23$. The solid, dashed, dotted, dash-dotted
  lines are for $g^{2}/\mu^{2}=0.0$, 3.0, 5.77, and 9.32 GeV$^{-2}$,
  respectively. The red data is $R_{1.4} = 11.0_{-0.6}^{+0.9}$ km,
  which is the radius of $1.4\, M_{\odot}$ constrained by the
  observations of GW170817 \citep{cap20}. The blue and green regions
  show the mass and radius estimates of PSR J0030+0451 derived from
  NICER data by Riley et al. \cite{ril19} ($R = 12.71_{-1.19}^{+1.14}$
  km, $M = 1.34_{-0.16}^{+0.15}\, M_{\odot}$) and Miller et
  al. \cite{mil19} ($R = 13.02_{-1.06}^{+1.24}$ km, $M
  =1.44_{-0.14}^{+0.15}\, M_{\odot}$).}
\label{fig1}
\end{figure}

\begin{figure*}[tbp]
\centering
\includegraphics[width=1.0\linewidth]{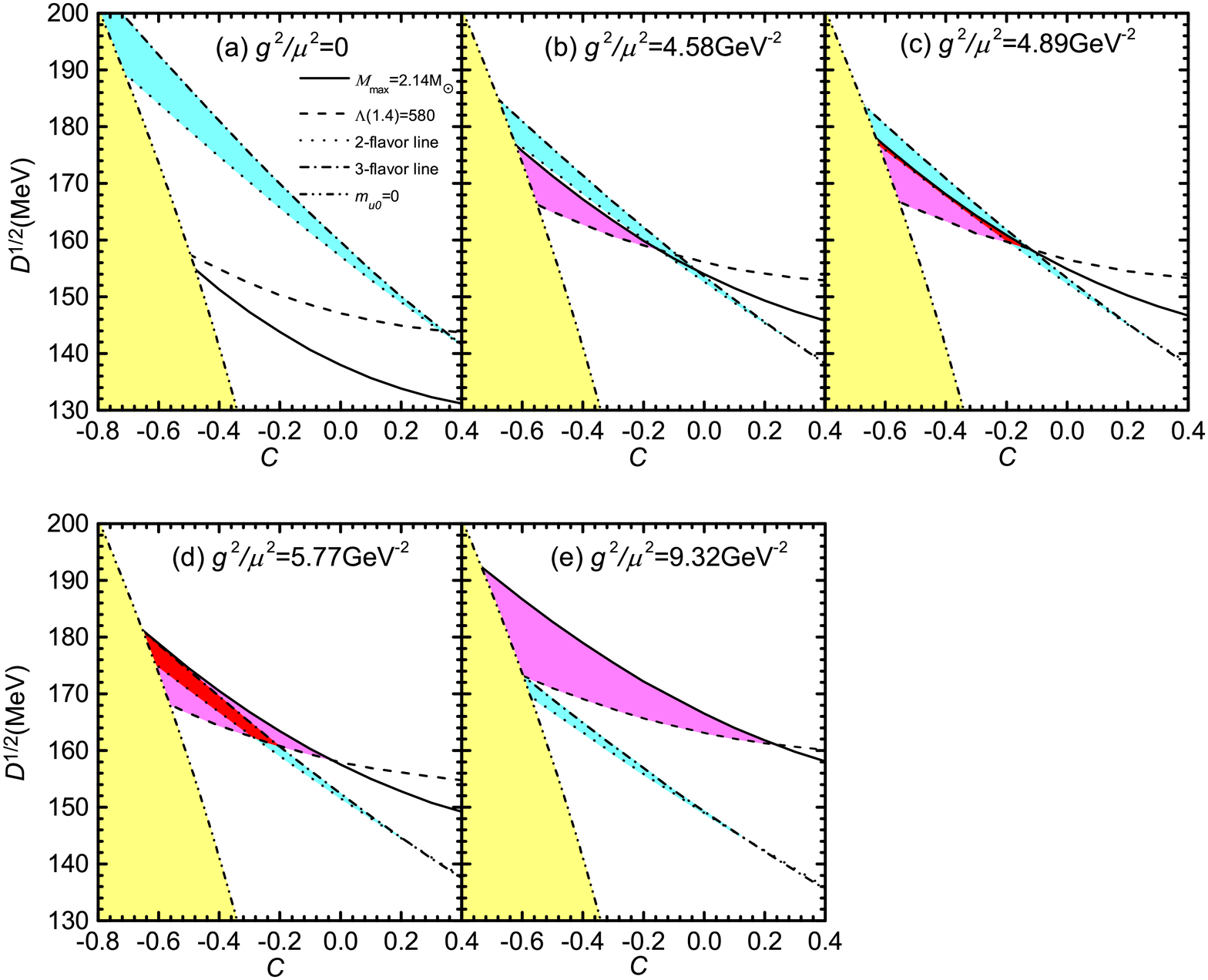}
\caption{ (Color online) Constraints on $D^{1/2}$ and $C$ for
  $g^{2}/\mu^{2}$ = 0 (a), $g^{2}/\mu^{2}$ = 4.58 GeV$^{-2}$ (b),
  $g^{2}/\mu^{2}$ = 4.89 GeV$^{-2}$ (c), $g^{2}/\mu^{2}$ =
  5.77 GeV$^{-2}$ (d), and $g^{2}/\mu^{2}$ = 9.32 GeV$^{-2}$ (e),
  respectively.  The red-shadowed regions in panels (c) and (d)
  indicate the allowed parameter spaces.  (See text for details.)}
\label{fig2}
\end{figure*}

\begin{figure*}[tbp]
\centering
\includegraphics[width=1.0\linewidth]{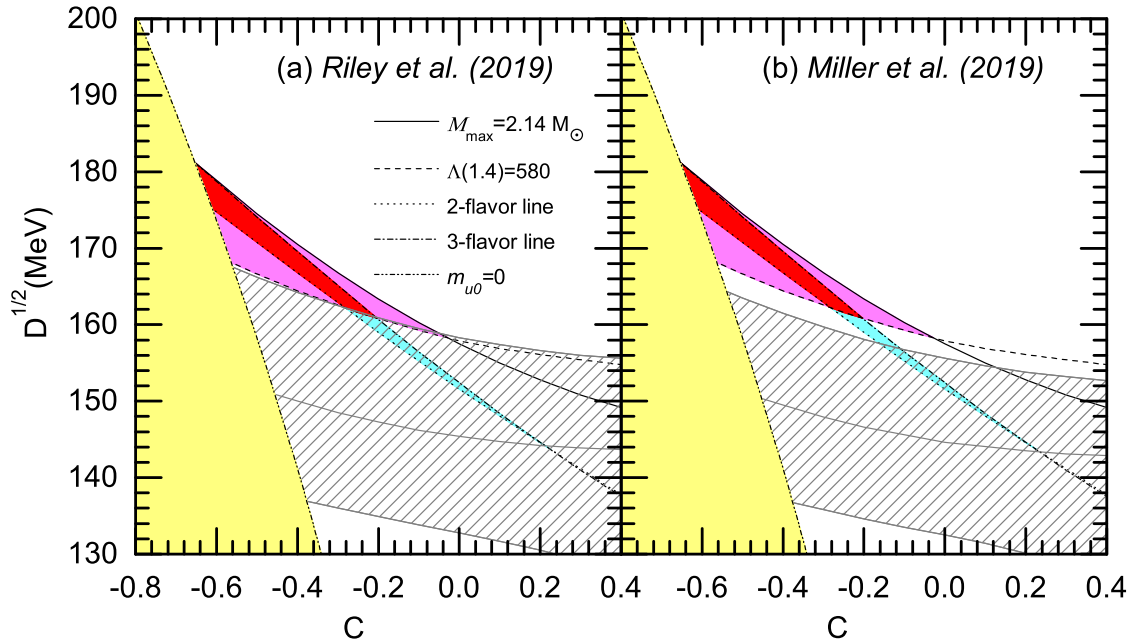}
\caption{ (Color online) Constraints on $D^{1/2}$ and $C$ for
  $g^{2}/\mu^{2}$ = 5.77 GeV$^{-2}$.  The gray-shaded regions in
  panels (a) and (b) indicate the parameter spaces restricted by the
  mass and radius of PSR J0030+0451 derived from the NICER observation
  by Riley et al. \citep{ril19} and Miller et al. \citep{mil19},
  respectively.}
\label{fig3}
\end{figure*}

The mass-radius relations of SSs for different non-Newtonian gravity
parameters are shown in Fig.\ \ref{fig1}. We choose $D^{1/2} = 161.3$
MeV, $C = -0.23$ because for this set of parameter, the observations
of PSR J0740+6620, GW170817 and PSR J0030+0451 (only for mass and
radius data given by Riley et al. \cite{ril19}) can be satisfied
simultaneously when the non-Newtonian gravity parameter
$g^{2}/\mu^{2}=5.77$, as will be shown in Fig.\ \ref{fig3}. The
dash-dotted line for $g^{2}/\mu^{2}=9.32$ satisfies the
constraints on
 PSR J0030+0451 set by NICER data and the radius data
derived by Capano et al. \citep{cap20}. The
corresponding set of parameter $D^{1/2} = 161.3$ MeV, $C = -0.23$, and
$g^{2}/\mu^{2}=9.32$ is ruled out by the constraints employed by this
paper later, which can be seen in Fig.\ \ref{fig2}(e).

We investigate the allowed parameter space of QMDD model according to
the following five constraints \citep[e.g.,][]
{sch97,wei11,wei12,pi15,zh18,yang20,bac20}:

First, as pointed out by Backes et al. \citep{bac20}, the quark masses
could become negative at high densities and a negative mass has no
physical meaning, resulting in a regime where the model is not valid.
Following Backes et al. \citep{bac20}, we present the invalid
$D^{1/2}$--$C$ parameter regions in Fig.\ \ref{fig2} (namely, the
yellow-shaded regions), which are separated with other areas by
requiring $m_{u0}=0$ at $n=1.5$ fm$^{-3}$ (around ten times the
nuclear saturation density).  These yellow-shaded regions are ruled
out because for the parameters located in these regions, $m_{u0}$
becomes negative on densities lower than $n=1.5$ fm$^{-3}$, which may
happen in the cores of the massive SSs.

Second, the existence of SSs is based on the idea that the presence of
strange quarks lowers the energy per baryon of a mixture of $u$, $d$
and $s$ quarks in beta equilibrium below the energy of the most stable
atomic nucleus, $^{56}$Fe ($E/A\sim 930$ MeV) \citep{wit84} \footnote{It
is common practice to compare the energy of SQM to $^{56}$Fe.
The energy per baryon of $^{56}$Fe, however, is only the third lowest after $^{62}$Ni and $^{58}$Fe.}. This
constraint results in the 3-flavor lines (the dash-dotted lines) shown
in Fig.\ \ref{fig2}.

Here we want to stress that atomic nuclei do not transition to (lumps of) SQM, and
neither the EOS of ordinary nuclear matter nor NN scattering data are impacted by the possible absolute stability of SQM.
The reason is that the creation of SQM requires a significant fraction of strange quarks to be present.
Conversion of an $^{56}$Fe nucleus, for instance, into SQM requires a very high-order weak interaction to simultaneously change dozens of $u$ and $d$ quarks into $s$ quarks. The probability of this happening is astronomically small. For lower baryon numbers, the conversion requires a lower-order weak interaction, but finite-size effects and the positive electrostatic potential of SQM destabilize small junks of SQM so that they become unstable even if SQM is stable in bulk.

The third constraint is given by assuming that non-strange quark
matter (i.e., two-flavor quark matter made of only $u$ and $d$ quarks)
in bulk has an energy per baryon higher than the one of $^{56}$Fe,
plus a 4 MeV correction coming from surface effects
\citep{far84,mad99,wei11,zh18}.  By imposing $E/A\geq 934$ MeV on
non-strange quark matter, one ensures that atomic nuclei do not
dissolve into their constituent quarks. This leads to the 2-flavor
lines (dotted lines) in Fig.\ \ref{fig2}.  The cyan-shaded areas
between the 3-flavor lines (the dash-dotted lines) and the 2-flavor
lines (dotted lines) in Fig.\ \ref{fig2} show the allowed
$D^{1/2}$--$C$ parameter regions where the second and the third
constraints described just above are fulfilled.

The fourth constraint is that the maximum mass of SSs must be greater
than the mass of PSR J0740+6620, $M_{\rm max} \geq 2.14\,
M_{\odot}$. By employing this constraint, the allowed parameter space
is limited to the region below the solid lines in Fig.\ \ref{fig2}.

The last constraint follows from $\Lambda(1.4)\leq 580$, where
$\Lambda(1.4)$ is the dimensionless tidal deformability of a
$1.4\, M_{\odot}$ star. The parameter space satisfies this constraint
corresponds to the region above the dashed lines in
Fig.\ \ref{fig2}. The magenta-shaded areas between
the solid lines and the dashed lines in
Fig.\ \ref{fig2} show the allowed $D^{1/2}$--$C$ parameter regions
where both constraints from the mass PSR J0740+6620 and the tidal deformability
of GW170817 are fulfilled.

By imposing all the five constraints discussed above, the allowed
$D^{1/2}$--$C$ parameter space of QMDD model is restricted to the
red-shadowed regions shown in Fig.\ \ref{fig2}(c) and \ref{fig2}(d),
which are obtained for non-Newtonian gravity parameter values of
$g^{2}/\mu^{2}=4.89$ GeV$^{-2}$, and $g^{2}/\mu^{2}=5.77$ GeV$^{-2}$,
respectively. An overlapping region where all the five constraints are
simultaneously satisfied does not exist for all other cases shown in
Fig.\ \ref{fig2}, panels (a), (b), (e), which correspond to
$g^{2}/\mu^{2}=0$, $g^{2}/\mu^{2}=4.58$ GeV$^{-2}$, and
$g^{2}/\mu^{2}=9.32$ GeV$^{-2}$, respectively.

From Fig.\ \ref{fig2}(a), one sees that for the case of
$g^{2}/\mu^{2}=0$, the five constraints mentioned above cannot be be
satisfied simultaneously.  This situation continues as the value of
$g^{2}/\mu^{2}$ becomes bigger until it is as large as 4.58
GeV$^{-2}$, in which case the $M_{\rm max}=2.14\, M_{\odot}$ line, the
$\Lambda(1.4)=580$ line and the 2-flavor line intersect at the point
(158.3, -0.15)(see Fig.\ \ref{fig2}(b)).  The allowed parameter space
vanished entirely for $g^{2}/\mu^{2} > 9.32$ GeV$^{-2}$, as shown in
Fig.\ \ref{fig2}(e).

Let us focus on Fig.\ \ref{fig2}, panels (b), (c) and (d)
once again. In Fig.\ \ref{fig2}(b), the $M_{\rm max}=2.14\, M_{\odot}$
line, the $\Lambda(1.4)=580$ line and the 2-flavor line intersect at
the point (158.3, $-0.15$), which means that the lower limit of
$D^{1/2}$ is 158.3 MeV.  In Fig.\ \ref{fig2}(c), the $M_{\rm
  max}=2.14\, M_{\odot}$ line, the $\Lambda(1.4)=580$ line and the
3-flavor line intersect at the point (158.5, $-0.12$), which means
that the upper limit of $C$ is $-0.12$. Whereas, in
Fig.\ \ref{fig2}(d) , the $M_{\rm max}=2.14\, M_{\odot}$ line, the
3-flavor line and the $m_{u0}$=0 line intersect at the point (181.2,
$-0.65$), which suggests that the upper limit of $D^{1/2}$ is 181.2
MeV and the lower limit of $C$ is $-0.65$.

In addition, one can see from Fig.\ \ref{fig2}(e) that the largest
allowed maximum mass for our SQM model that can satisfy all the above
five constraints simultaneously is reached at $D^{1/2} = 173.1$ MeV,
$C = -0.60$ and $g^{2}/\mu^{2} = 9.32$ GeV$^{-2}$, which is $2.42 \,
M_{\odot}$.

Recently, the NICER observations of the isolated pulsar PSR J0030+0451
produced two independent measurements of the pulsar's mass and
equatorial radius: $M = 1.34_{-0.16}^{+0.15}\, M_{\odot}$ and $R_{\rm eq}
= 12.71_{-1.19}^{+1.14}$ km \citep{ril19}, and $M
=1.44_{-0.14}^{+0.15}\, M_{\odot}$ and $R_{\rm eq} =
13.02_{-1.06}^{+1.24}$ km \citep{mil19}. In Fig.\ \ref{fig3}, these
data on the $M$--$R$ plane is translated into the $D^{1/2}$--$C$ space
(namely, the gray-shaded regions) for the case of $g^{2}/\mu^{2}$ =
5.77 GeV$^{-2}$. The gray lines in Fig.\ \ref{fig3}(a) are for
($M\, (M_{\odot}$), $R$\,(km)) sets (1.49, 11.52), (1.34, 12.71), and
(1.18, 13.85) from top to bottom, and these parameter sets correspond
to the data given by Riley et al. \citep{ril19}. The gray lines
in Fig.\ \ref{fig3}(b) are for (1.59, 11.96), (1.44, 13.02), and
(1.30, 14.26) from top to bottom, and these parameter sets come from
the data given by Miller et al.\ \citep{mil19}.

We can see from Fig.\ \ref{fig3}(a) that the allowed parameter
space constrained by the mass and radius of PSR J0030+0451 given by
Riley et al.\ \citep{ril19} (the gray-shaded region) marginally
overlaps with the allowed region (the red-shadowed region) restricted
by the five constraints mentioned earlier.  However, for the
observational data given by Miller et al.\ \citep{mil19} in
Fig.\ \ref{fig3}(b), no such overlapping region exist. This means that
there exists a very tiny allowed parameter space for
  which our SQM model agrees with the observations related to
PSR J0740+6620, GW170817 and PSR J0030+0451 simultaneously if one
employs the mass and radius given by Riley et al.\ \citep{ril19}. On
the other hand, if the data from Miller et al. \citep{mil19} is
employed, these observations cannot be explained
simultaneously. Although we only show the case of $g^{2}/\mu^{2}$ =
5.77 GeV$^{-2}$ in Fig.\ \ref{fig3}, we have checked some other cases
between 4.58 GeV$^{-2} < g^{2}/\mu^{2} < $ 9.32 GeV$^{-2}$ and find
that one always arrives at the above conclusion.

\section{summary}\label{Sec VI}

In this paper, we have investigated the effects of non-Newtonian
gravity on the properties of SSs and constraint the parameter space of
the QMDD model using astrophysical observations related to PSR
J0740+6620 and GW170817.  Similarly to the results presented in
Ref.\ \citep{yang20}, we found that these observations cannot be
explained by the SQM model employed in this paper if the non-Newtonian
gravity effects are not included.  In other words, the existence of
SSs is ruled out in this case.

\begin{figure*}[tbp]
	\centering
	\includegraphics[width=1.0\linewidth]{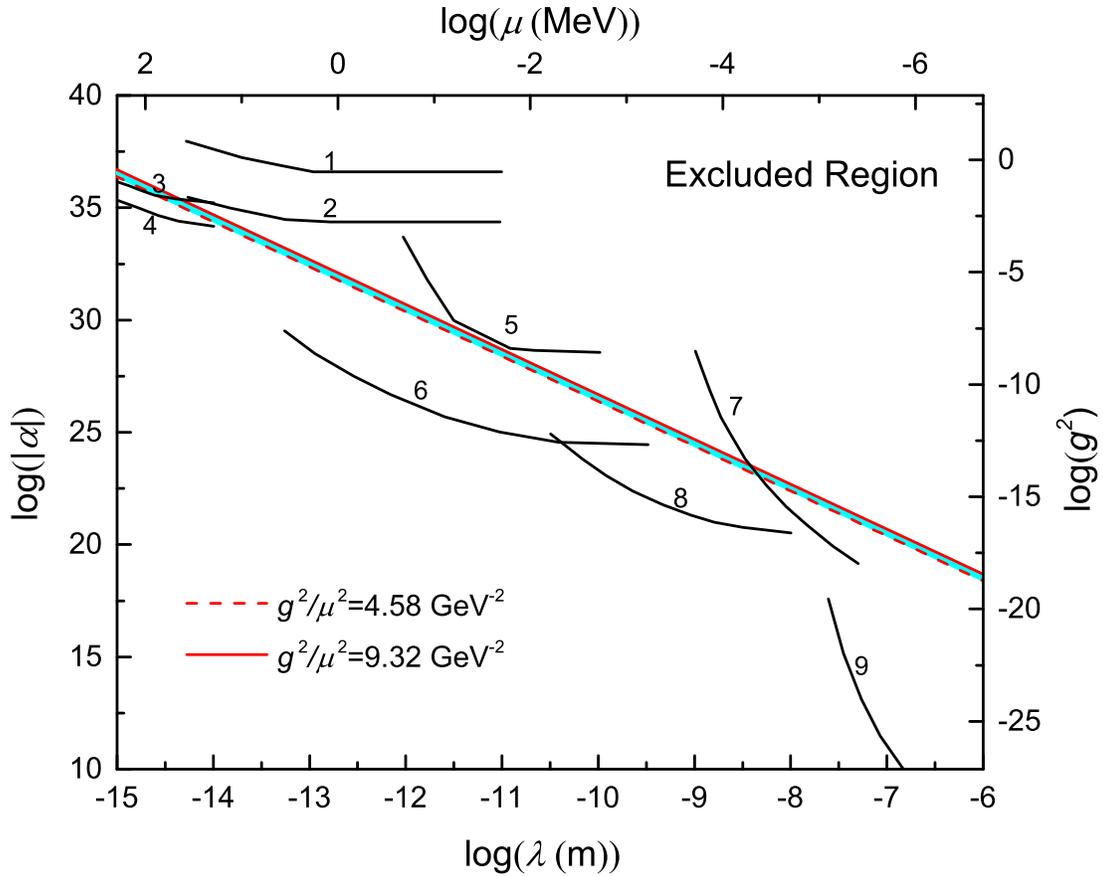}
	\caption{(Color online) Upper bounds on the strength parameter
          $|\alpha|$ respectively the boson-nucleon coupling constant
          $g$ as a function of the range of the Yukawa force $\mu$
          (bottom) and the mass of hypothetical bosons (top), set by
          different experiments as used in Ref. \citep{yang20}: curves
          1 and 2 refer to constraints from np scattering of scalar
          and vector bosons, respectively \citep{Kamyshkov08}; 3 and 4
          are constraints extracted from charge radii and binding
          energies of atomic nuclei, respectively \citep{Xu13}; 5 was
          established from the spectroscopy of antiprotonic He atoms
          and 6 from neutron total cross section scattering of
          $^{208}{\rm Pb}$ nuclei \citep{Pokotilovski06}; 7 is from an
          experiment measuring the Casimir force between a Au-coated
          microsphere and a silicon carbide plate
          \citep{Klimchitskaya20}; 8 is obtained by measuring the
          angular distribution of 5~\AA ~neutrons scattered off of an
          atomic xenon gas \citep{Kamiya15}; 9 shows the constraints
          from the force measurements between a test mass and rotating
          source masses of gold and silicon \citep{Chen16}. The
          cyan-shaded strip corresponds to 4.58 GeV$^{-2}\leq
          g^{2}/\mu^{2}\leq$ 9.32 GeV$^{-2}$. } \label{fig4}
\end{figure*}

Considering the non-Newtonian gravity effects, for the current quark
mass $m_{u0}=2.16$ MeV, $m_{d0}=4.67$ MeV, and $m_{s0}=93$ MeV
\citep{zyla20}, an allowed parameter space of $D^{1/2}$ and $C$ exists
only when 4.58 GeV$^{-2}\leq g^{2}/\mu^{2}\leq$ 9.32 GeV$^{-2}$, and
the parameters of the QMDD model are restricted to 158.3 MeV$\leq
D^{1/2}\leq$ 181.2 MeV and $-0.65 \leq C \leq -0.12$. As shown in
Fig.\ \ref{fig4}, theoretical bounds on $g^{2}/\mu^{2}$ of 4.58
GeV$^{-2}\leq g^{2}/\mu^{2}\leq$ 9.32 GeV$^{-2}$ for which QSs are
found to exist (indicated by the cyan-colored strip in the figure) is
excluded by some experiments (curves labeled 4, 6, 8, 9) but allowed
by others (curves labeled 1, 2, 5 and parts of curves 3 and 7).

We also find that the largest allowed maximum mass of SSs for the QMDD
model is $2.42 \, M_{\odot}$, corresponding to the parameter set
$D^{1/2} = 173.1$ MeV, $C = -0.60$ and $g^{2}/\mu^{2} = 9.32$
GeV$^{-2}$. Therefore, even considering the non-Newtonian effect, the
GW190814's secondary component with mass $2.59_{-0.09}^{+0.08}\,
M_{\odot}$ \cite{abb20} could not be a static SS. However, it could be
a rigid or differentially rotating SS \citep{zhou19}.

Moreover, by translating the mass and radius of PSR J0030+0451
observed by NICER into the $D^{1/2}$--$C$ space, we find that for the
analysis by Riley et al.\ \citep{ril19}, there exists a very tiny
allowed parameter space for which SSs constructed with
the QMDD model agree with the observations related to PSR
J0740+6620, GW170817 and PSR J0030+0451 simultaneously; but for the
analysis by Miller et al.\ \citep{mil19}, these observations cannot be
explained simultaneously.

\acknowledgements The authors are especially indebted to the anonymous referee for his/her valuable comments.
We thank J. Schaffner-Bielich for discussions on the stability of strange quark matter.
This work is supported by National SKA Program of China No. 2020SKA0120300, and the Scientific Research
Program of the National Natural Science Foundation of China (NSFC,
grant Nos.\ 12033001, 11773011, and 11447012). F.W.\ is supported
through the U.S.\ National Science Foundation under Grants PHY-1714068
and PHY-2012152.

\end{document}